\documentclass[pre,aps,twocolumn,superscriptaddress]{revtex4-1}
\usepackage{amsmath, amsthm, amssymb,float}
\usepackage{amsfonts}
\usepackage{graphicx}
\usepackage{dcolumn}
\usepackage{bm}
\usepackage{braket}
\usepackage{relsize}
\usepackage{textcomp}
\usepackage{cases}
\usepackage[normalem]{ulem}
\usepackage[titletoc,title]{appendix}
\usepackage{cancel}


\usepackage[colorlinks=true, urlcolor=blue, anchorcolor=blue, citecolor=blue,filecolor=blue,linkcolor=blue,menucolor=blue,pagecolor=blue]{hyperref}

\usepackage{textcomp}

\begin{document}

\title{A giant disparity and a dynamical phase transition in large deviations of the time-averaged size of stochastic populations}
\author{Pini Zilber}
\email{pizilber@gmail.com}
\author{Naftali R. Smith}
\email{naftali.smith@mail.huji.ac.il}
\author{Baruch Meerson}
\email{meerson@mail.huji.ac.il}
\affiliation{Racah Institute of Physics, Hebrew University of
Jerusalem, Jerusalem 91904, Israel}

\begin{abstract}
We study large deviations of the time-averaged size of stochastic populations
described by a continuous-time Markov jump process. When the expected population
size $N$ in the steady state is large, the large deviation function (LDF) of the time-averaged population size can be evaluated by using a WKB (after Wentzel, Kramers and Brillouin) method,  applied directly to the master equation for the Markov process.  For a class of models that we identify, the direct WKB method predicts a giant disparity between the probabilities of observing an unusually small and an unusually large values of the time-averaged population size. The disparity results from a qualitative change in the ``optimal" trajectory of the underlying classical mechanics problem. The direct WKB method also predicts, in the limit of $N\to \infty$, a singularity of the LDF, which
can be interpreted as a second-order dynamical phase transition. The
transition is smoothed at finite $N$, but the giant disparity remains. The smoothing effect is captured by the van-Kampen system size expansion of the exact master equation near the attracting fixed point of the underlying deterministic model. We describe the giant disparity at finite $N$ by developing a different variant of WKB method, which is applied in conjunction with the Donsker-Varadhan
large-deviation formalism and involves subleading-order calculations in $1/N$.

\end{abstract}
\maketitle

\nopagebreak

\section{Introduction}
\label{sec_intro}

Stochastic population models describe a multitude of processes in nature and society \cite{Delbrueck,Bartlett,Karlin,Bailey,Nisbet,Andersson,Allen,Tsimring,Allen2015}.
Examples include reactions among particles, dynamics of biological populations, spread of
diseases, inventions, opinions and rumors, dynamics of financial markets, and many other
processes. A convenient mathematical description of stochastic populations is provided by  Markov jump processes
\cite{Kampen,Gardiner}. An important class of these processes has a nontrivial steady
state, where the gain and loss processes balance each other on average. A standard characterization
of a fluctuating steady state is in terms of its steady-state probability distribution. For a single population in a steady state this distribution, $\pi_n$, describes
the probabilities to observe $n = 0, 1, 2,  \dots$ individuals, which we will call particles.

In recent years there has been a growing interest in a different characterization of fluctuating steady states: by
the probability distributions of \emph{long-time averages} of observables, see Refs. \cite{Touchette2009,Touchette2018} and
references therein. In the context of stochastic populations a natural quantity of this type is the long-time average of the (fluctuating) population size. For many stochastic systems, the probability of observing any value of
the time-averaged size, which is different from the expected one, becomes exponentially small
as the averaging time $T$ increases.

A similar property, and the ensuing large deviation function
(LDF) that encodes it, have received much recent attention from physicists in the context
of continuous diffusion processes, described by Langevin equations \cite{Touchette2009,Touchette2018,TouchettePRL2018,Lecomte}.
Recently this characterization has been extended \cite{MeersonZilber} to one of the best known jump
processes in physics:
the Ehrenfest Urn Model \cite{EUM}. The basic version of the
Ehrenfest Urn Model involves two urns with $N$ identical balls. The balls hop randomly, one ball at a time, from one urn to the other. For non-interacting balls the LDF of the long-time average of the number of balls in a given urn can be calculated
exactly \cite{MeersonZilber} by using the Donsker-Varadhan (DV) large-deviation formalism \cite{DonskerVaradhan}, which we briefly review below. If the balls  interact, the DV
formalism can still be used numerically \cite{MeersonZilber}, but it does not allow for analytical solution. However, when the total number of
balls $N$ is very large, one can circumvent the DV method and employ instead a WKB approximation, directly applied to the master equation \cite{Kubo,Gang,Dykman,AssafMeerson}. In this way  one obtains a controlled large-$N$ approximation
of the LDF \cite{MeersonZilber}.

Here we follow this line of work and
identify a class of stochastic populations, for which the direct
WKB method predicts a giant disparity between the probabilities of observing an unusually small and an unusually large values of the time-averaged population size. This disparity results from a qualitative change of the character of the optimal (least-action) trajectory of an underlying Hamiltonian classical mechanics problem. The direct WKB method also predicts
a jump in the second derivative of the LDF, which can
be interpreted as a second-order dynamical phase transition. The sharp transition is smoothed by
finite-$N$ effects, unaccounted for by the leading-order WKB approximation, but the giant disparity remains. We show it by developing a different version of WKB theory, which involves subleading-order WKB calculations and
is used in conjunction with the DV large-deviation formalism.

Now let us be more specific. Consider a jump process which admits a nontrivial steady state, and denote the population size (the number of particles) in a specific realization of the process  at time $t$ by $n\left(t\right)$. We are interested in the probability, $\mathcal{P}(\bar{n}_T = Na)$, that the time averaged population size $\bar{n}_T$, defined as
\begin{equation}\label{eq_nbarDefinition}
\bar{n}_T = \frac{1}{T}\int_0^T n\left(t\right)dt ,
\end{equation}
is equal to $Na$, where $N\gg 1$ is the expected population size, and $a$ is a real non-negative number. When $T$ is very large,  $\mathcal{P}(\bar{n}_T = Na)$ satisfies the large deviations principle, which states that any deviation from the expected value is exponentially unlikely to occur:
\begin{equation}\label{eq_LDP}
-\ln \mathcal{P}(\bar{n}_T= Na) \simeq T I\left(a,N\right) .
\end{equation}
The rate function $I\left(a,N\right)$ is the focus of our interest.  It is a non-negative function of $a$ that vanishes at its most probable value  which, for $N\gg 1$,  is $a\simeq 1$. In the limit $T\rightarrow \infty$, $I\left(a,N\right)$ is independent of the initial condition.  An analytical calculation of $I\left(a,N\right)$ is in general impossible.  When $N\gg 1$, one can employ a direct WKB approximation \cite{Kubo,Gang,Dykman,AssafMeerson}, as it has been recently done for the Ehrenfest Urn Model \cite{MeersonZilber}. In this approximation
the $a$- and $N$-dependences of the rate function $I\left(a,N\right)$ factorize \cite{MeersonZilber}.
In the class of models that we will work with here, the rate function $I\left(a,N\right)$, as predicted by the direct WKB approximation, is proportional to $N$:
\begin{equation}\label{eq_Nscalinglinear}
I\left(a,N\right)\simeq N f(a),\quad \quad N \gg 1.
\end{equation}
It is natural to call the function $f(a)$ the \emph{intensive} rate function.

For many stochastic populations the intensive rate function $f(a)$ is an analytic function of $a$, and the Ehrenfest Urn Model provides one such example \cite{MeersonZilber}. As we will see shortly, another such example is provided by the immigration-death process
$\emptyset \rightleftarrows A$, a simple and well studied reversible jump process. However, as we find here, there is a class of population models where
$f(a)$, as described by the direct WKB method,  vanishes on the whole interval $0\leq a\leq1$. This corresponds to a giant disparity between the probabilities of observing an unusually small and unusually large values of $a$. Furthermore, the predicted $f(a)$ is non-analytic at $a=1$. We will identify this class of models and exemplify it by still another simple reversible process: the branching-coalescence process $A \rightleftarrows 2A$. Then we will obtain more accurate results for this model, by going beyond the direct WKB method.

Here is a plan of the remainder of the paper. In Sec.~\ref{sec_WKB0_formalism} we outline the direct WKB method and consider two simple examples: the immigration-death process and the branching-coalescence process. We show that the intensive rate function $f(a)$ is analytic in the former case, and non-analytic and vanishing at $0\leq a\leq1$ in the latter.
In Sec.~\ref{sec_WKB1}  we apply WKB approximation, in the leading and subleading orders with respect to $1/N$, to the DV method. In the leading order this calculation reproduces the results of the direct WKB method. In the subleading order the $f(a)=0$ degeneracy is removed due to finite-$1/N$ effects. We also compute the probability distribution $\pi_n(a)$ of the population size, conditional on a specified value of $a$. We show that, in the phase-transition region of $a$, the  distribution $\pi_n(a)$ is bimodal, which reflects the character of the optimal trajectories of the system in this region. We discuss our results and some possible generalizations in Sec. \ref{sec_discussion}.

\section{Direct WKB method}
\label{sec_WKB0_formalism}

\subsection{General}
\label{general}

We start this section by considering a general single-step jump process for a single population. A generalization to multiple-step processes is immediate \cite{Dykman,EscuderoK,AM2010}. The process is governed by the master equation \cite{Kampen,Gardiner}
\begin{align}\label{eq_singleJumpMaster}
\partial_t P_n\left(t\right) &= W_+(n-1)P_{n-1}(t) + W_-(n+1)P_{n+1}(t) \nonumber \\
& - \left[ W_+(n) + W_-(n) \right] P_n\left(t\right),
\end{align}
for the evolution of the probability $P_n\left(t\right)$ of observing $n$ particles at time $t$. The first two terms on the right describe the incoming rates into state $n$ from the states $n-1$ and $n+1$, whereas the last term describes the total outgoing rate from state $n$. We assume throughout this work that the birth and death rates $W_+(n)$ and $W_-(n)$ are such that Eq.~(\ref{eq_singleJumpMaster}) admits a nontrivial steady-state solution. We will also assume that
the birth and death rates are disparate so that the expected population size $N$ in the steady state is large, and exploit the large parameter $N\gg 1$.
Let us expand the process rates $W_\pm$ in the powers of $N$. Introducing the rescaled population size $q=n/N$, we obtain \cite{EscuderoK,AM2010}
\begin{equation}\label{eq_ratesExpansion}
W_\pm(n) = W_\pm(Nq) = N w_\pm(q) + u_\pm(q) + ... ,
\end{equation}
where $w_\pm(q)$ and $u_\pm(q)$ are assumed to be $O(1)$.

If one multiplies  the master equation \eqref{eq_singleJumpMaster}  by $n$, sums it up over all possible values of $n$ and ignores fluctuations \cite{Kampen,Gardiner},  one arrives at the \emph{deterministic rate equation}
\begin{equation}\label{rateeq}
\frac{dq}{dt}= w_+(q)- w_-(q).
\end{equation}
The stationary population, residing in a vicinity of $n=N$, corresponds to an attracting fixed point of Eq.~(\ref{rateeq}) at $q=1$, so that  $w_+(1)-w_-(1)=0$ and $w_+'(1)-w_-'(1)<0$, where prime denotes the $q$-derivative.

The direct WKB method for an analysis of fluctuations \cite{Kubo,Gang,Dykman,AssafMeerson} exploits the large parameter $N\gg 1$ by applying the WKB ansatz
\begin{equation}\label{eq_WKBansatz}
P_n\left(t\right) = e^{-N s\left(\frac{n}{N},t \right)}= e^{-N s\left(q,t \right)}
\end{equation}
directly to the master equation. Additionally, we assume $n\gg 1$, treat $q = n/N$ as a continuous variable, and evaluate the action $s$ at $\frac{n\pm 1}{N} = q \pm \frac{1}{N}$ by Taylor expanding it around $q$ \cite{Kubo,Gang,Dykman}:
\begin{equation}
\label{eq_actionTaylor}
s \! \left( \! q\pm\frac{1}{N},t \! \right) \! = \! s \! \left(q,t\right)\pm\frac{1}{N}\frac{\partial s \! \left(q,t\right)}{\partial q}+\frac{1}{2N^{2}}\frac{\partial^{2}s \! \left(q,t\right)}{\partial q^{2}}+\dots.
\end{equation}
Using this expansion of $s$ in the WKB ansatz \eqref{eq_WKBansatz} and plugging the ansatz into the master equation \eqref{eq_singleJumpMaster}, we arrive at an equation which involves $q$, $\partial_q s$ and $\partial_t s$ in different orders of $1/N$, and treat it order by order. In the leading order this procedure yields a Hamilton-Jacobi equation for the ``action" $s(q,t)$ \cite{Kubo,Gang,Dykman}:
\begin{equation}\label{eq_HJ}
\frac{\partial s}{\partial t} + H_0\left( q, \frac{\partial s}{\partial q} \right) = 0,
\end{equation}
where
\begin{equation}\label{eq_singleJumpH0}
H_{0}(q,p)=w_{+}(q)(e^{p}-1)+w_{-}(q)\left(e^{-p}-1\right)
\end{equation}
is the model-specific Hamiltonian. As one can see, the population size $q$ plays the role of the coordinate of an effective ``particle", whereas $p =\partial_q s$ is the momentum. The action along a trajectory $q(t),p(t)$ is given by
\begin{equation}\label{eq_action}
s\left(a,T\right)=\int_{0}^{T}p\left(t\right)\dot{q}\left(t\right)dt-ET,
\end{equation}
where $E=H_{0}\left[q\left(t\right),p\left(t\right)\right]$ is the (conserved) energy of the ``particle". The constraint  $n_T = Na\,$ ($a\geq 0$) can be accounted for by adding a term $-\mu\int_{0}^{T}q\left(t\right)dt$, where $\mu$ is a  Lagrange multiplier, to the action. In the Lagrangian formulation the constrained Lagrangian
is
\begin{equation}\label{constrainedL}
L(q,\dot{q}) =L_0(q,\dot{q}) - \mu q ,
\end{equation}
where $L_0(q,\dot{q})$ is the unconstrained Lagrangian, corresponding to the unconstrained Hamiltonian $H_0(q,p)$. The constrained Hamiltonian is equal to \cite{LLMechanics}
\begin{equation}\label{constrainedHfirst}
H(q,p)=p \dot{q}-L(q,\dot{q})=p \dot{q}-L_0(q,\dot{q})+\mu q,
\end{equation}
where $\dot{q}$ should be expressed through $q$ and $p$ from the relation $p=\partial L/\partial{\dot{q}}$. As a result,
the constrained Hamiltonian takes the form \cite{MeersonZilber}
\begin{equation}\label{H}
H(q, p) = H_0(q,p)+\mu q .
\end{equation}
As one can see, the constraint is enforced via a simple additive term $\mu q$  in the Hamiltonian. This simplicity occurs because $\partial L/\partial{\dot{q}}$ does not depend on $\mu$, see Eq.~(\ref{constrainedL}).
The phase-space  trajectories $q(t)$ and $p(t)$ satisfy the Hamilton's equations
\begin{eqnarray}
  \dot{q} &=& \frac{\partial H}{\partial p} = \frac{\partial H_0}{\partial p}, \label{qdot}\\
  \dot{p} &=& -\frac{\partial H}{\partial q} = -\frac{\partial H_0}{\partial q} - \mu. \label{pdot}
\end{eqnarray}
Among all such trajectories,  we must choose the one which satisfies appropriate boundary conditions and the condition
\begin{equation}\label{averageq}
\frac{1}{T}\int_0^{T} q(t) dt = a,
\end{equation}
which ultimately sets the value of $\mu$. If there is more than one such trajectory, we  must choose the ``optimal" one which minimizes the action \eqref{eq_action}. Having found it, we can evaluate $-\ln\mathcal{P}\left(\bar{n}_{T}=Na\right)\simeq Ns\left(a,T\right)$, where $s\left(a,T\right)$ is its corresponding action.

In the Hamiltonian description the attracting fixed point $q=1$ of the deterministic rate equation (\ref{rateeq}) becomes a saddle point $(q=1,\,p=0)$ of the unconstrained Hamiltonian $H_0$. For the constrained Hamiltonian (\ref{H})  this saddle point is shifted. For finite $T$ the action $s\left(a,T\right)$ depends on the initial condition $q\left(t=0\right)$, and must be minimized with respect to the final condition $q\left(t=T\right)$. As $T$ increases, the effective ``particle" spends more and more time in a close vicinity of the above-mentioned shifted saddle point of the constrained Hamiltonian \cite{MeersonZilber}. For $T\rightarrow \infty$ this saddle point dominates both the action (\ref{eq_action}) (where the first term vanishes) and the integral over time in Eq.~(\ref{averageq}). Then, in view of the Eq.~(\ref{averageq}), the $q$-value of this fixed point must be equal to $a$. The corresponding value of $p=p(a)$ can be obtained from the algebraic equation
\begin{equation}\label{pofa}
\frac{\partial H_0(a,p)}{\partial p}=0 ,
\end{equation}
which follows from Eq.~(\ref{qdot}). At this stage we are already constraining the process by $a$, so the Lagrange multiplier $\mu$ becomes unnecessary. The optimal action, evaluated at the saddle point, is very simple:
\begin{equation}\label{eq_activationAction}
s\left(a,T\right)=-ET=-H_{0}\left[a,p\left(a\right)\right]T.
\end{equation}
Finally, combining Eqs.~(\ref{eq_LDP}), (\ref{eq_WKBansatz}) and (\ref{eq_activationAction}), we obtain the intensive rate function $f(a)$, defined by Eq.~(\ref{eq_Nscalinglinear}):
\begin{equation}\label{eq_fEqualsH0}
f(a) =\frac{s\left(a,T\right)}{T} = -H_{0}\left[a,p\left(a\right)\right].
\end{equation}
Equation~(\ref{eq_fEqualsH0}) has quite a general form, and it also holds for multi-step processes. Let us specialize Eq.~(\ref{eq_fEqualsH0})  to the single-step process (\ref{eq_singleJumpMaster}). Using Eqs.~(\ref{eq_singleJumpH0}) and (\ref{pofa}),
we obtain the value of $p$ at the fixed point:
\begin{equation}\label{psinglestep}
p=\frac{1}{2}\ln  \frac{w_-(a)}{w_+(a)}.
\end{equation}
As a result, Eq.~(\ref{eq_fEqualsH0}) yields
\begin{equation}\label{eq_fSingleJump}
f(a) = \left[ \sqrt{w_+(a)} - \sqrt{w_-(a)} \right]^2 ,
\end{equation}
so one only needs to know the leading-order terms of the birth and death rates~(\ref{eq_ratesExpansion}).
Equation~(\ref{eq_fSingleJump}) reproduces the recent result \cite{MeersonZilber} for the rate function of the time-averaged number of balls in one of the urns in the Ehrenfest Urn Model.

\subsection{$0 \rightleftarrows A$}
\label{0toA}

As another simple example, consider the immigration-death process
\begin{equation}\label{eq_0AA0process}
\emptyset \xrightarrow{N} A,\quad A \xrightarrow{1} \emptyset
\end{equation}
with the per-capita immigration rate $N$ and the per-capita death rate $1$. The process rates are $W_+(n) = N$ and $W_-(n) = n$. This simple process obeys the detailed balance condition
\begin{align}\label{eq_detailed_balance}
\pi_n W_+(n) = \pi_{n+1}W_-(n+1),
\end{align}
where the equilibrium distribution $\pi_n$  is Poisson:
\begin{align}\label{eq_poisson_immigrationDeath}
\pi_n = \frac{e^{-N}N^n}{n!},\quad n=0,1,\dots.
\end{align}

\begin{figure}[h]
\includegraphics[width=0.33\textwidth,clip=]{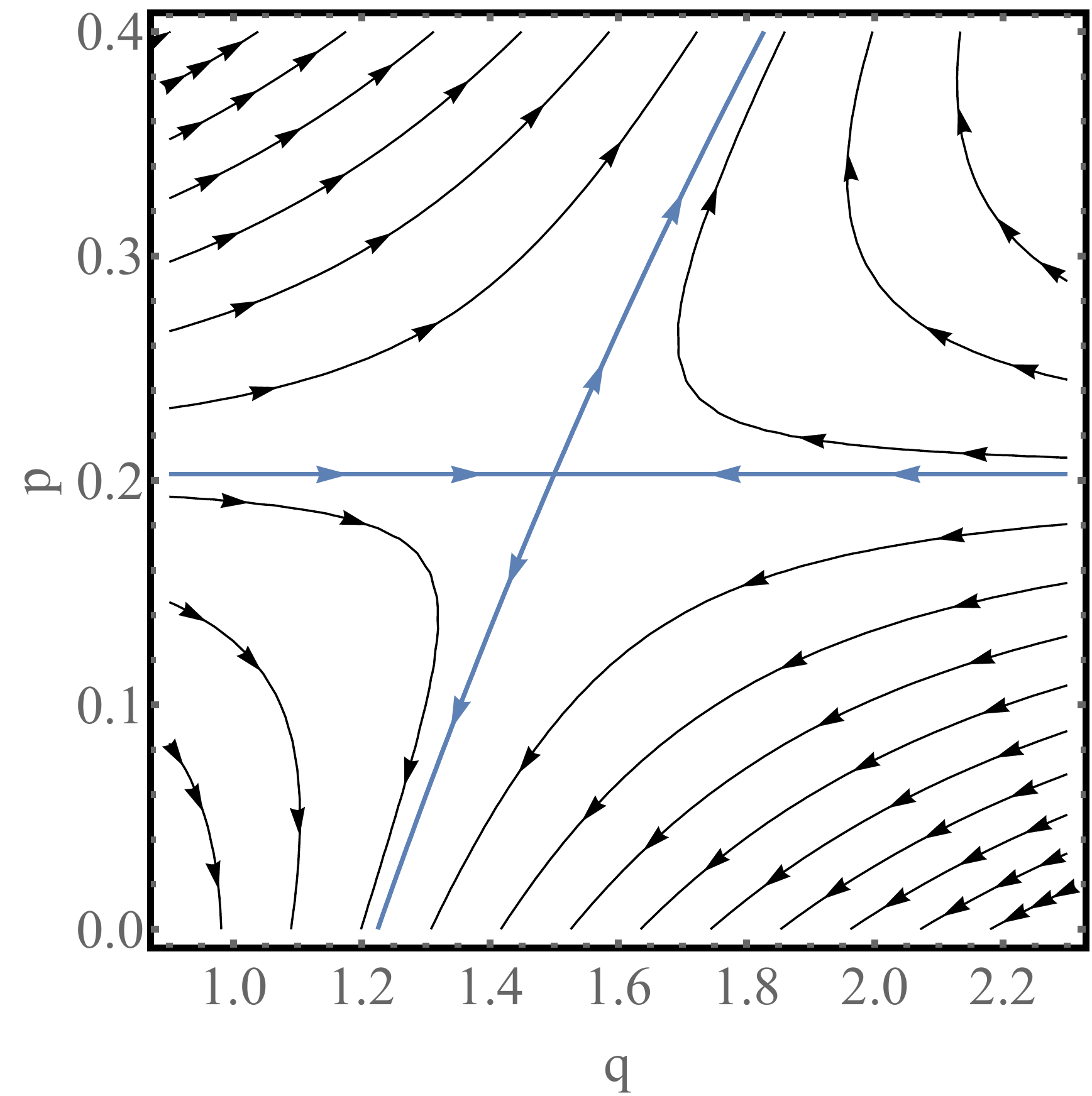}
\caption{The phase portrait of the constrained Hamiltonian (\ref{Himdeath}) of the immigration-death process $\emptyset\rightleftarrows A$ for $\mu=1-\sqrt{2/3}$. The saddle point is at $(q,p)=[3/2, (1/2) \ln(3/2)]$.}
\label{fig_0AA0_contours0}
\end{figure}

Now we assume $N\gg 1$ and $n\gg 1$ and turn to the WKB description. We immediately note that the leading-order rates $w_+(q) = 1$ and $w_-(q) = q$ are exact here. Figure \ref{fig_0AA0_contours0} shows the phase portrait of the constrained Hamiltonian,
\begin{equation}\label{Himdeath}
H(q,p,\mu)=e^p-1+q(e^{-p}-1)+\mu q .
\end{equation}
The shifted saddle point is at $\left(q,p\right)=  \left[\left(1-\mu\right)^{-2}\!,-\ln\left(1-\mu\right)\right]$. It exists  for $-\infty<\mu<1$, and its coordinate $q$ can take any value from zero to infinity.

\begin{figure}[h]
\includegraphics[width=0.4\textwidth,clip=]{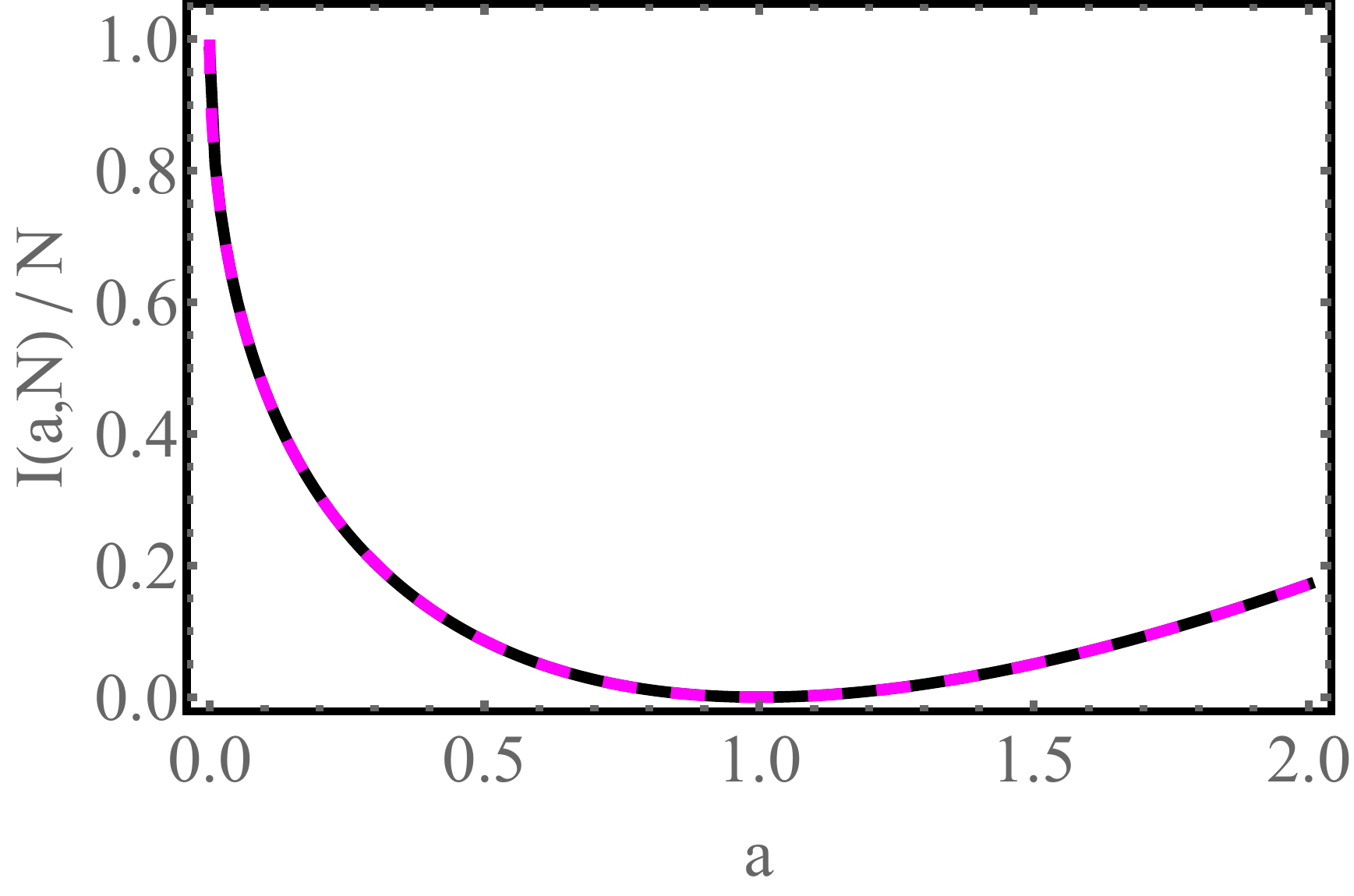}
\caption{The rescaled rate function $I\left(a,N\right)/N=\left(1-\sqrt{a}\right)^{2}$  for the immigration-death process $\emptyset\rightleftarrows A$ (dashed line) is compared with numerical results obtained by the DV method \cite{DonskerVaradhan} for $N=10$ (solid line).}
\label{fig_0AA0_WKB0}
\end{figure}

Plugging the rates $w_+(q) = 1$ and $w_-(q) = q$ into Eq. (\ref{eq_fSingleJump}), we obtain the intensive rate function of the long-time average population size in this system:
\begin{equation}\label{eq_0AA0f}
f(a) = \left(1-\sqrt{a}\right)^{2} .
\end{equation}
The function $f(a)$ is strictly convex. It vanishes at the expected value $a=1$ and is positive for all other $a$. It is shown in Fig. \ref{fig_0AA0_WKB0} alongside with numerical results, obtained by the DV method
\cite{DonskerVaradhan}, which we briefly discuss in the beginning of Sec. \ref{sec_WKB1}.
The perfect agreement, even for a moderate value of $N=10$, is explained by the fact that, for this simple process, the WKB result (\ref{eq_0AA0f}) is exact. The WKB result is also exact for the Ehrenfest Urn Model with non-interacting balls \cite{MeersonZilber}.

\subsection{$A\rightleftarrows 2A$: Giant disparity of probability and dynamical phase transition at $N\to \infty$}
\label{sec_WKB0_examples}

Expressions (\ref{eq_fEqualsH0}) and, for single-step processes,~(\ref{eq_fSingleJump}) are not always correct. Indeed, the shifted saddle point with coordinate $a$ may not exist, or it may be unaccessible for some initial conditions. But even if it exists and is accessible, there can be a different trajectory  with a lesser action, which does not stay in a close vicinity of the shifted saddle point. We will illustrate these general arguments on the example of a well-known branching-coalescence process
\begin{equation}\label{eq_A2A2AAprocess}
A \xrightarrow{1} 2A,\quad 2A \xrightarrow{2/N} A,
\end{equation}
with per-capita rates $1$ and $2/N$, respectively. This process exhibits nontrivial dynamics only if it starts from a nonzero number of particles, and we will suppose that this is the case. The branching rate $W_+(n)$ is given by the product of the per-capita branching rate $1$ and the current population size $n$. The coalescence rate $W_-(n)$ is given by the product of the per-capita coalescence rate $2/N$ and the number of \emph{pairs} in the population of $n$ particles,  $n\left(n-1\right)/2$. As a result,
\begin{equation}\label{eq_A2A2AArates}
W_+(n) = n, \quad W_-(n) = \frac{n\left(n-1\right)}{N}.
\end{equation}
This process also obeys the detailed balance condition~(\ref{eq_detailed_balance}), and the equilibrium distribution is again Poisson:
\begin{align}
\label{Poisson2}
\pi_{n}=\frac{N^{n}}{\left(e^{N}-1\right)n!},\quad n=1,2,\dots.
\end{align}

\begin{figure}[h]
\includegraphics[width=0.4\textwidth,clip=]{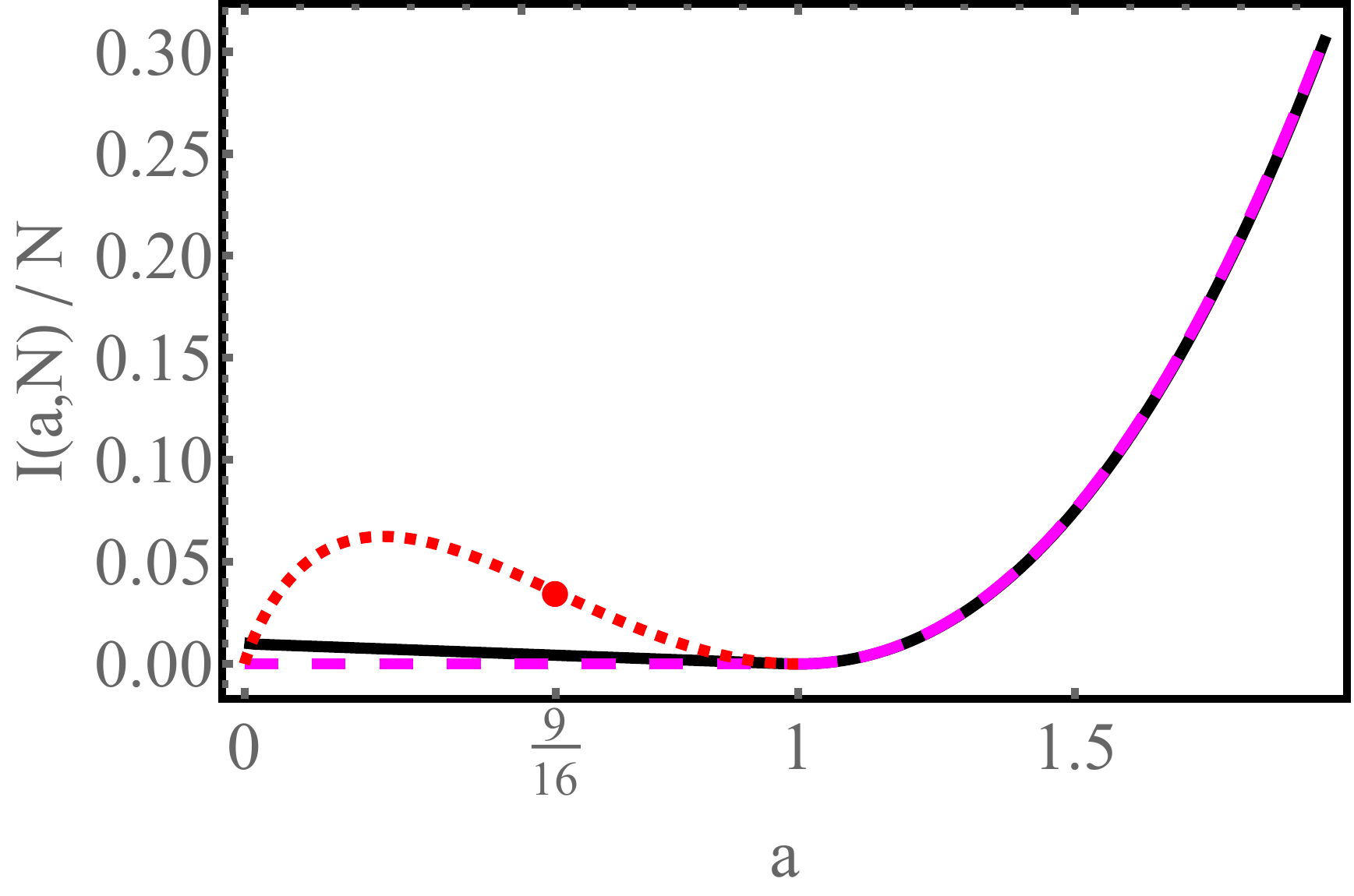}
\caption{The rescaled rate function  $I\left(a,N\right)/N$ for the branching-coalescence process $A\rightleftarrows 2A$: theory versus numerics. Dotted line: prediction of Eq.~(\ref{eq_A2A2AA_incorrectf}), which turns out to be correct only for $a\geq 1$. Dashed line: correct leading-order result from Eq.~(\ref{eq_A2A2AA_correctf}). Solid line: numerical results for $I\left(a,N\right)/N$, obtained by the DV method \cite{DonskerVaradhan} for $N=100$.}
\label{fig_A2A2AA_WKB0}
\end{figure}

\begin{figure}
\includegraphics[width=0.22\textwidth,clip=]{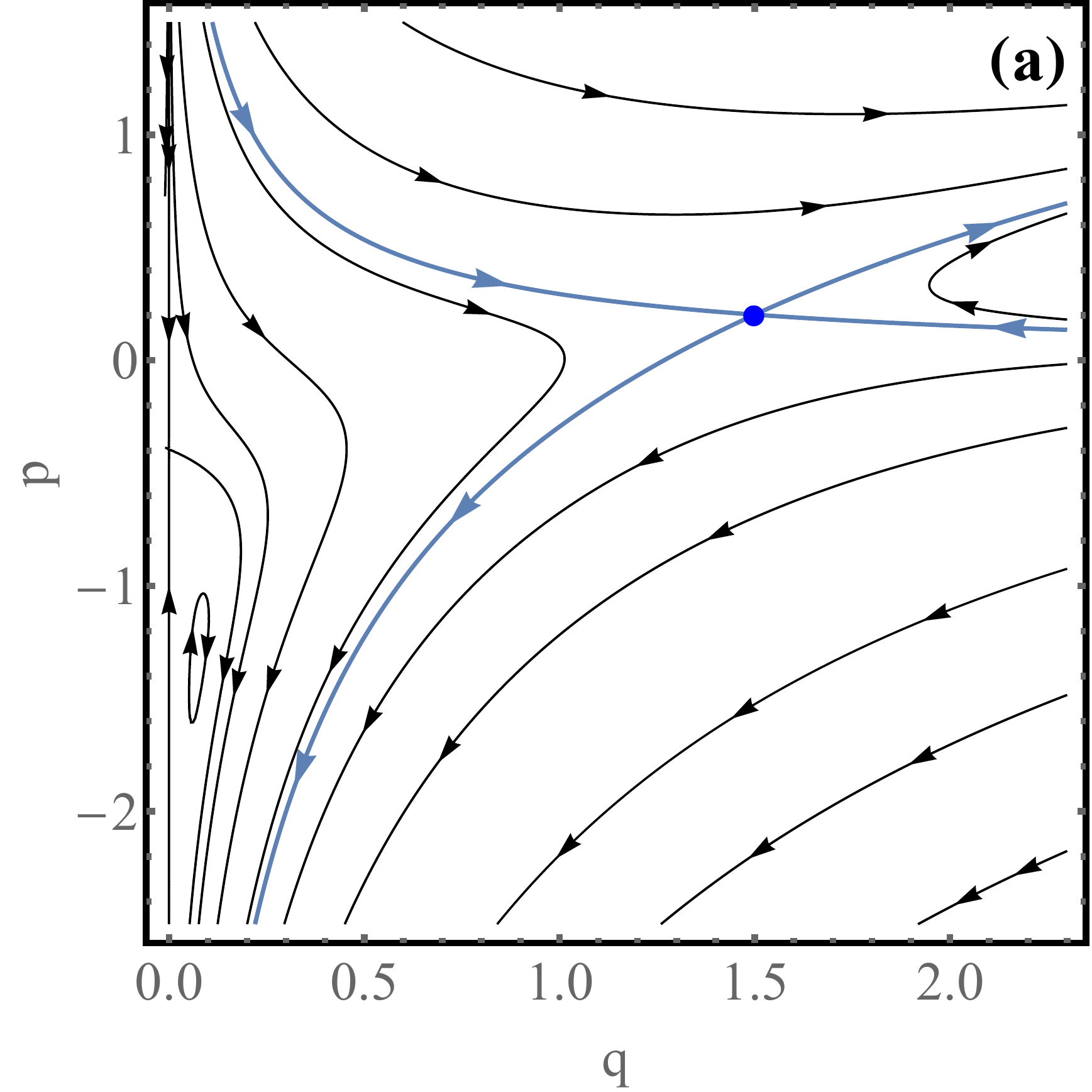}
\includegraphics[width=0.23\textwidth,clip=]{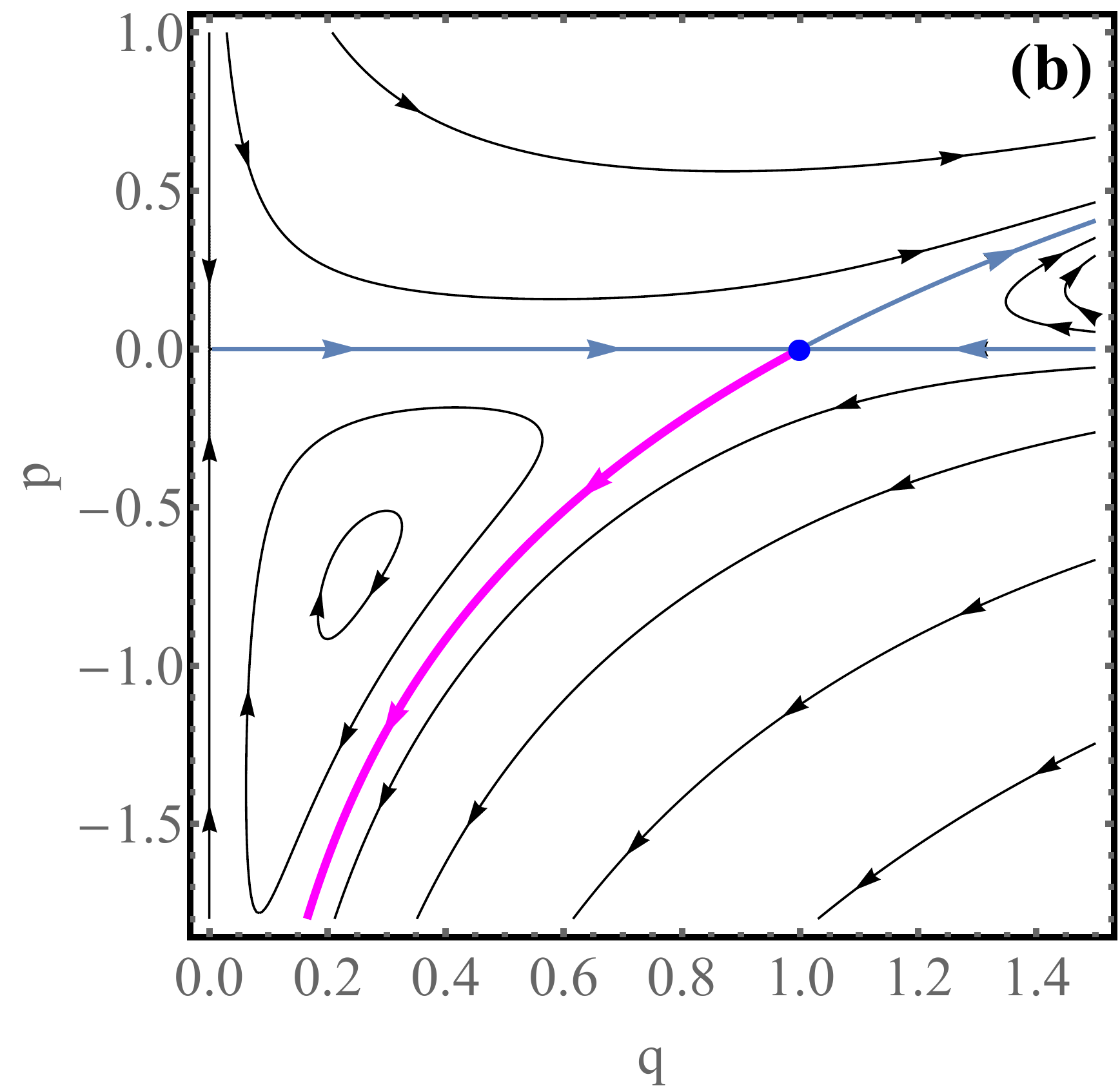}
\includegraphics[width=0.23\textwidth,clip=]{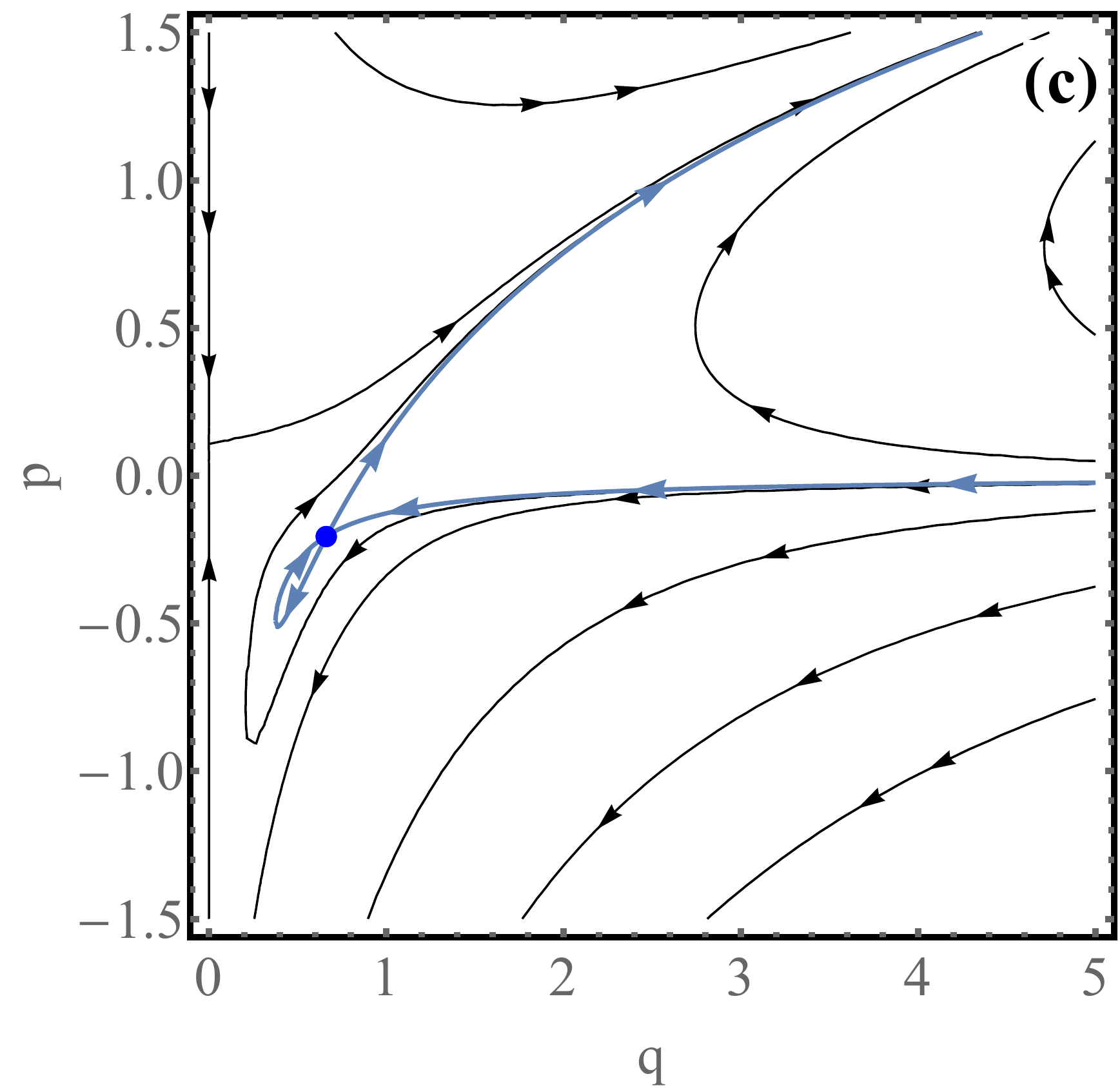}
\includegraphics[width=0.23\textwidth,clip=]{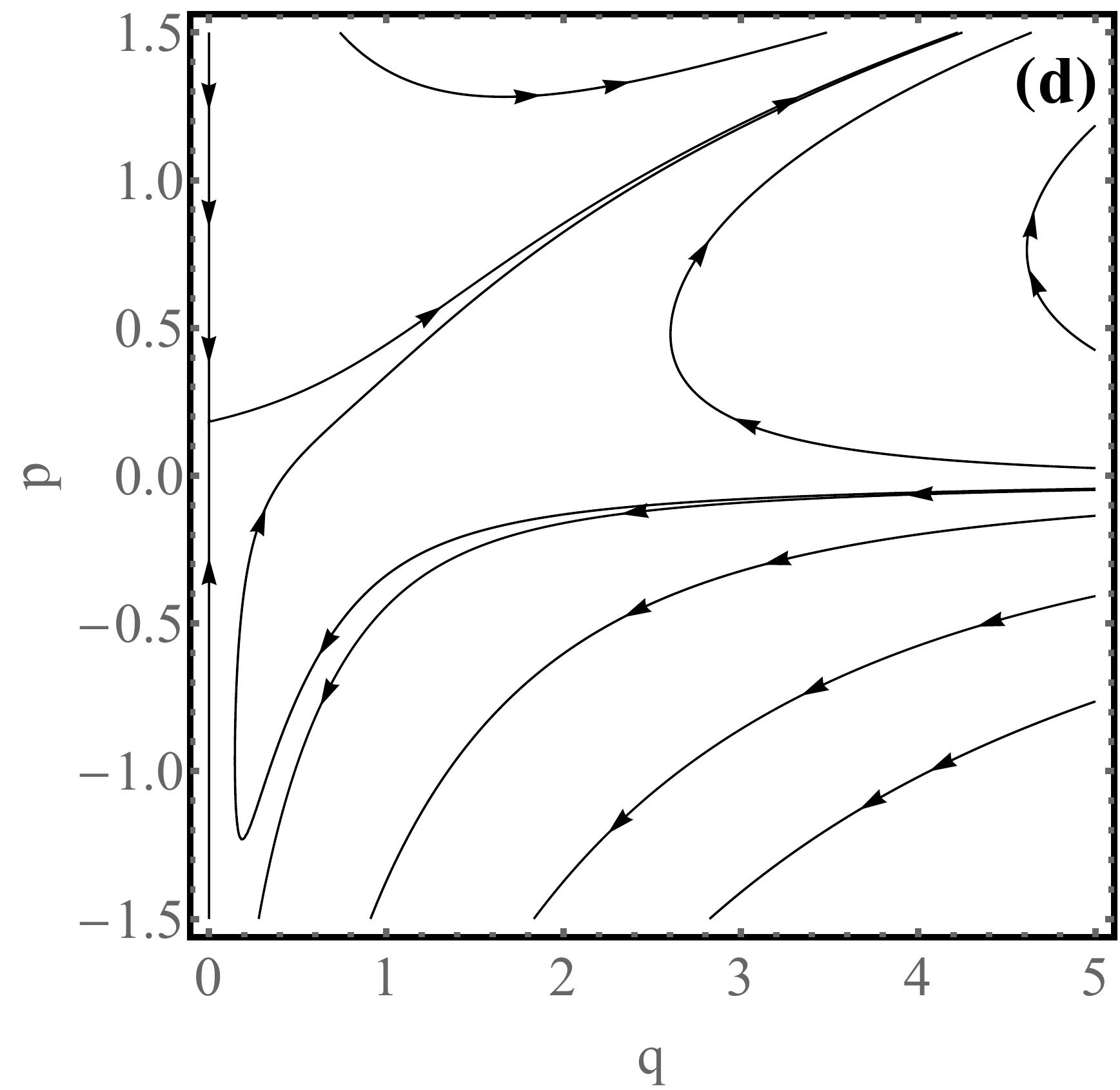}
\caption{The phase portrait of the constrained Hamiltonian~(\ref{H0Ato2Amu}) of the branching-coalescence process $A\rightleftarrows 2A$ for $\mu=0.32...$ (a), $0$ (no constraint) (b), $-0.12...$ (c), and $-0.2$ (d). The $q$-values of the nontrivial saddle point (shown by the fat dot) in panels (a), (b) and (c), are $3/2$, $1$, and $2/3$, respectively. The thick trajectory on panel (b) corresponds to the family of instanton solutions~(\ref{instanton}).}
\label{constrainedH}
\end{figure}

However, as we will see now, the large deviation function of the long-time average population size is more interesting here than for the immigration-death process. Applying the direct WKB method and  using Eqs.~\eqref{eq_ratesExpansion} and \eqref{eq_A2A2AArates}, we identify the leading-order rates $w_+(q) = q$ and $w_-(q) = q^2$. Recklessly using Eq.~(\ref{eq_fSingleJump}), we would obtain the intensive rate function
\begin{equation}\label{eq_A2A2AA_incorrectf}
f(a) = \left(\sqrt{a}-a\right)^{2} ,
\end{equation}
shown in Fig.~\ref{fig_A2A2AA_WKB0} by the dotted line.  This function is obviously non-convex for $a < 9/16$. Actually, it is non-convex, according to the supporting line definition of convexity, for all $0<a<1$.
This arises suspicion, and indeed Eq.~(\ref{eq_A2A2AA_incorrectf}) turns out to be incorrect on the whole interval $0< a<1$.
In order to see why, let us explore the phase portrait of the system. The deterministic equation (\ref{rateeq}) takes the form $\dot{q}=q-q^2$. The attracting fixed point is again $q=1$, but now there is also a repelling fixed point at $q=0$. This fact turns out to be crucial, as the effective mechanical system can now stay for a long time at the fixed point $q=0$ without the need to ``spend" any action.  Let us see how it works. The phase portraits of the unconstrained Hamiltonian,
\begin{equation}\label{H0Ato2A}
H_{0}\left(q,p\right)=q\left(e^{p}-1\right)+q^{2}\left(e^{-p}-1\right),
\end{equation}
is shown in Fig.~\ref{constrainedH}b. The repelling and attracting fixed points of the deterministic dynamics become saddle points $(q,p)=(0,0)$ and $(q,p)=(1,0)$, respectively. In addition, an elliptic fixed point appears at $(q,p)=(1/4, -\ln 2)$.

For the flow described by the constrained Hamiltonian,
\begin{equation}\label{H0Ato2Amu}
H\left(q,p,\mu\right)=q\left(e^{p}-1\right)+q^{2}\left(e^{-p}-1\right)+\mu q,
\end{equation}
the three fixed points become $\mu$-dependent. The saddle point $(0,0)$ of the unconstrained Hamiltonian becomes $[0, \ln(1-\mu)]$. It exists for $\mu<1$. The nontrivial saddle point exists for $\mu>-1/8$, and the elliptic point exists for $-1/8<\mu<1$. At $\mu=-1/8$ the latter two fixed points merge, and for $\mu<-1/8$ they disappear: compare panels \textit{c} and \textit{d} in
Fig.~\ref{constrainedH}. As one can check, for all $\mu\geq -1/8$, the value of $q$ of the nontrivial saddle point is greater than or equal to
$9/16$; the equality occurs at $\mu=-1/8$. The absence of a nontrivial saddle point with $q<9/16$ (see Fig.~\ref{constrainedH}d) explains why Eq.~(\ref{eq_A2A2AA_incorrectf}) cannot be correct for $a<9/16$.

What happens for $9/16<a<1$? Here the nontrivial saddle point is accessible for some initial conditions, and inaccessible for others. More importantly,  there is a trajectory of a different type, which corresponds to $\mu=0$,  provides \emph{any} specified value of $a\in (0,1)$ and has a lesser action. This trajectory is closely related to the exact one-parameter family of instanton solutions of the unconstrained  Hamilton equations:
\begin{equation}\label{instanton}
q\left(t,C\right)=\frac{1}{1+e^{t+C}},\quad p\left(t,C\right)=-\ln\left(1+e^{t+C}\right).
\end{equation}

\begin{figure}
\includegraphics[width=0.38\textwidth,clip=]{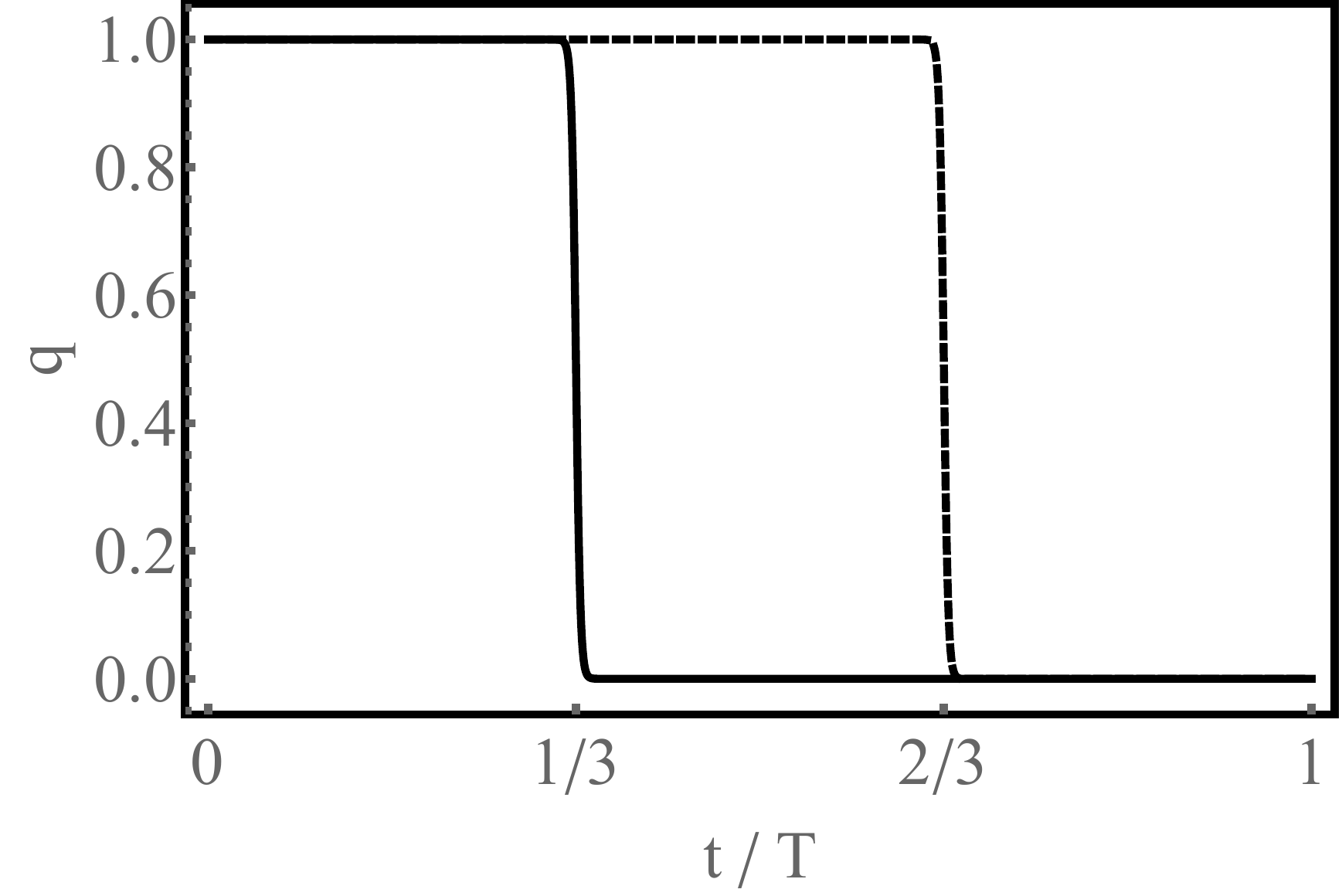}
\caption{Two examples of the instanton solutions~\eqref{instanton} satisfying condition \eqref{averageq} for $a\leq 1$ up to exponentially small corrections in $T\gg 1$: $q\left(t,C=-T/3\right)$ (solid line, corresponds to $a=1/3$) and $q\left(t,C=-2T/3\right)$ (dashed line, corresponds to $a=2/3$).}
\label{zeroEnergySolutions}
\end{figure}

These solutions, depicted in Fig.~\ref{zeroEnergySolutions}, satisfy the boundary conditions $q\left(t=-\infty,C\right)=1$, $q\left(t=\infty,C\right)=0$, $p\left(t=-\infty,C\right)=0$ and $p\left(t=\infty,C\right)=-\infty$. They correspond to the zero-energy phase trajectory $p=\ln q$, indicated by the thick line on Fig.~\ref{constrainedH}b. When $T\gg 1$, they can serve as approximate solutions on the finite interval $0<t<T$, and they are accurate up to corrections which are exponentially small with respect to $T\gg 1$. Furthermore, one can always choose the constant $C$ so that condition (\ref{averageq}) is satisfied for any $0<a<1$. The action (\ref{eq_action}) for each of these trajectories is equal (again, up to exponentially small corrections in $T$) to
\begin{equation}\label{saless1}
s = \int_1^0 p(q)\,dq= \int_1^0 \ln q\,dq = 1.
\end{equation}
Crucially, this action is independent of $T$. As a result, the intensive rate function $f(a)=s/T$ \emph{vanishes} at $T\to \infty$ and $0<a<1$, and we arrive at
\begin{equation} \label{eq_A2A2AA_correctf}
f\left(a\right)=\begin{cases}
0, & a\leq1,\\
\left(a-\sqrt{a}\right)^{2}, & a\geq1.
\end{cases}
\end{equation}
Therefore, the probabilities of observing $a<1$ and $a>1$ are dramatically different. Notably,
$f(a)$ is continuous with its first derivative $f'(a)$ at $a=1$. The second derivative, however,  has a jump at $a=1$ which can be interpreted as a dynamical phase transition of second order. The appearance of a giant disparity in $f(a)$, and of a second-order phase transition, in a simple reversible one-population model is quite remarkable.

The sharp transition, however, is observed only in the limit $N\to \infty$.  A valuable insight into finite-$N$ effects can be achieved if we go beyond the WKB method and try to interpret
the unexpected result $f(a) = 0$ at $0\leq a\leq 1$ in the language of the exact microscopic process $A\rightleftarrows 2A$. We can easily do it for one particular value of $n$: $n=1$.
What is the probability to observe, for very large $T$, the average population size equal to $1$ \cite{fn_nmin}?
The only stochastic trajectory that can contribute to this average population size is the one where $n\left(t\right)=1$, while the branching process $A\to 2A$ is suppressed for the whole time $T\gg 1$.
The probability of the stochastic trajectory $n\left(t\right)=1$ during the time $T$ is $e^{-T}$, hence $I\left(a=1/N,N\right)=1$. Consequently, in the limit $N\gg 1$, we get $f\left(a=0\right)=0$. On the other hand, $I\left(a,N\right)$ vanishes at $a\simeq 1$ (for $N\gg 1$), so that $f\left(a=1\right)=0$. As we expect the rate function to be convex, it must satisfy $f\left(0\leq a\leq1\right) = 0$, in agreement with the first line of Eq.~(\ref{eq_A2A2AA_correctf}), predicted by the direct WKB method.

The direct-WKB prediction~(\ref{eq_A2A2AA_correctf}) for $f(a)$ is presented in Fig.~\ref{fig_A2A2AA_WKB0} alongside with numerical results  obtained by the DV method \cite{DonskerVaradhan} for $N=100$.
One can see excellent agreement for $a>1$ (which is not surprising, as in this regime the solution comes from the shifted saddle point). The numerical result at $a=1/N$ perfectly agrees with the exact result $I\left(a=1/N,N\right)/N = 1/N$, quoted above.
It follows immediately from $I\left(a=1/N,N\right)=1$, $I\left(a\simeq1,N\right)=0$ and the expected convexity of $I(a)$ that
\begin{equation}
\label{upper_bound}
\frac{I\left(a,N\right)}{N}\le\frac{1-a}{N}+O\left(\frac{1}{N^{2}}\right),\quad0\le a\le1.
\end{equation}
The numerics indeed indicate a linear behavior of  $I\left(a,N\right)/N$ as a function of $a$, with slope $-1/N$, so that the upper bound~(\ref{upper_bound}) is in fact very close to the actual value of $I\left(a,N\right)/N$.
Our next goal will be to derive a subleading WKB asymptotic which describes the finite $N$ effects observed in the numerics.

\section{WKB for DV}
\label{sec_WKB1}

\subsection{DV method}
\label{general1}

The exact DV method \cite{DonskerVaradhan} reduces the determination of the rate function $I\left(a,N\right)$ in the $T\rightarrow \infty$ limit to the solution of an eigenvalue problem. In most cases this eigenvalue problem can only be solved numerically. Figures~\ref{fig_0AA0_WKB0} and~\ref{fig_A2A2AA_WKB0} show such numerical solutions which we obtained for the processes $0\rightleftarrows A$ and  $A\rightleftarrows 2A$, respectively.  In this section we will show how to solve this eigenvalue problem analytically, albeit approximately, by exploiting the large parameter $N\gg 1$. We will do it by developing a different version of WKB theory, which is applied to the DV eigenvalue problem.

We will start with a brief description of the DV method. Consider a general jump process, governed by a master equation of the form
\begin{equation}\label{eq_masterWithL}
\partial_t P_n\left(t\right) = \sum_{m=1}^\infty L^+_{n,m}P_m(t).
\end{equation}
For a single-step process (\ref{eq_singleJumpMaster}) $\hat{L}^+$ is a tridiagonal matrix. The G\"{a}rtner-Ellis \cite{GartnerEllis} theorem states that the rate function $I\left(a,N\right)$ is given by the Legendre-Fenchel transform,
\begin{equation}\label{eq_LFtransform}
I\left(a,N\right) = \max_k\left\{kNa-\lambda\left(k,N\right)\right\},
\end{equation}
of the scaled cumulant generating function $\lambda\left(k\right)$, defined as
\begin{equation}\label{eq_SCGF_def}
\lambda\left(k,N\right)=\lim_{T\to\infty}\frac{1}{T} \ln \left\langle e^{TkNa}\right\rangle ,
\end{equation}
where $\braket{...}$ denotes averaging over possible values of $a$.
According to DV, $\lambda\left(k,N\right)$ is equal to the largest eigenvalue of an auxiliary operator $\hat{L}^k$:
\begin{equation}\label{eq_SCGF_byDV}
\lambda\left(k,N\right) = \xi_\text{max}\left(\hat{L}^k\right) ,
\end{equation}
where
\begin{equation}\label{eq_tiltedL}
L^k_{n,m} = L_{n,m} + nk\delta_{n,m}
\end{equation}
is a tilted version of the operator $\hat{L}$, which is the Hermitian conjugate of $\hat{L}^+$. For any value of $a$, the Legendre-Fenchel transform (\ref{eq_LFtransform}) associates a value of $k$.  We will  denote it by $k^{*}\left(a,N\right)$, so that
\begin{align}\label{eq_LF_kstar}
I\left(a,N\right)=k^{*}\left(a,N\right)Na-\lambda\left[k^{*}\left(a,N\right),N\right].
\end{align}
We used (a truncated version of) this method to perform our numerical calculations.

\subsection{$1/N$ expansion and WKB for DV}
\label{WKB1a}

Employing the small parameter $1/N$, we can seek the rate function perturbatively:
\begin{equation}\label{eq_IexpansionToSecondOrder_full}
I\left(a,N\right) = NI_0(a) + I_1(a) + O\left(\frac{1}{N}\right) .
\end{equation}
In the absence of the phase transition in the limit of $N\to \infty$,  both $I_0(a)$ and $I_1(a)$ are of order $1$.
For the branching-coalescence process (and other models which exhibits a phase transition of this type) $I_0(a)$ and $I_1(a)$ are of order $1$ only for $a>1$. For $a<1$ we should expect $I_0(a)=0$,  so the rate function is
\begin{equation}\label{eq_IexpansionToSecondOrder}
I\left(a,N\right) = I_1(a) + O\left(\frac{1}{N}\right),\quad a<1.
\end{equation}
It is this, more interesting case, that we will focus on \cite{lessinteresting}. We additionally assume that $a \gg 1/N$, as the WKB approximation does not hold for $n=O(1)$. As we will see, the WKB approximation also breaks down for $a$ very close to $1$, and we will use a different approximation there.

By virtue of Eqs.~\eqref{eq_LF_kstar} and~\eqref{eq_IexpansionToSecondOrder}, $k^{*}\left(a,N\right)$ must be of order $1/N$, and $\lambda\left[k^{*}\left(a,N\right),N\right]$ must be of order $1$, so we should set
\begin{align}\label{eq_k_expansionToSecondOrder}
k^{*}\left(a,N\right) &\simeq \frac{k_1^*(a)}{N} , \\
\label{eq_lambda_expansionToSecondOrder}
\lambda\left[k^{*}\left(a,N\right),N\right] &\simeq \lambda\left[\frac{k_1^*(a)}{N}, N\right] = \Lambda\left[k_{1}^{*}\left(a\right)\right] ,
\end{align}
where the unknown functions $k_1^*(a)$ and $\Lambda(k_1)$ are of order 1. The last equality in Eq.~(\ref{eq_lambda_expansionToSecondOrder}), where we introduce a new function $\Lambda\left[k_{1}^{*}\left(a\right)\right]$, follows from the demand that $\lambda\left[k_1^*(a)/N, N\right] = O(1)$ does not depend on $N$.
Plugging Eqs.~(\ref{eq_k_expansionToSecondOrder}) and~(\ref{eq_lambda_expansionToSecondOrder}) into Eq. (\ref{eq_LFtransform}) yields
\begin{equation}\label{eq_LFtransform_secondOrder}
I_1(a) = \max_{k_1}\left\{k_1a-\Lambda\left(k_1\right)\right\} .
\end{equation}
Let us determine the function $\Lambda(k_1)$, specializing our calculations to the branching-coalescence model. The auxiliary operator $\hat{L}^k$ of the DV method is
\begin{align}\label{eq_tiltedL_explicit}
L^{k}_{n,m} &= \frac{n\left(n-1\right)}{N}\delta_{n-1,m} + n\,\delta_{n+1,m} \nonumber \\
&- \left[n + \frac{n\left(n-1\right)}{N} - \frac{n}{N}k_1\right]\delta_{n,m} ,
\end{align}
where we have set $k \simeq k_1/N$, see Eq.~(\ref{eq_k_expansionToSecondOrder}). Let us consider the eigenvalue problem for the right eigenvector $r_n$,
\begin{align}\label{eq_eigenvaluesProblem}
\sum_{m=1}^{\infty} L^{k}_{n,m}r_m = \xi r_n,
\end{align}
and find the maximal eigenvalue $\xi =N\xi_0 + \xi_1+ \dots$ by using WKB approximation. As in the direct WKB method, we define $q=n/N$, make a WKB ansatz
\begin{align}
\label{WKBrn}
r_n \rightarrow r(q) = e^{NS(q)}
\end{align}
and treat $q$ as a continuous variable, so that
\begin{align}
r_{n\pm 1} \rightarrow r\left(q \pm \frac{1}{N}\right) &= e^{NS(q) \pm S'(q) + \frac{1}{2N}S''(q) + ...} .
\end{align}
When we plug this ansatz into Eq.~\eqref{eq_eigenvaluesProblem}, the factor $e^{NS(q)}$ will cancel out, so we will ignore it from now on. Before making the ansatz, we expand $S(q)$ in the powers of $1/N$:
\begin{align}\label{eq_ordersOfS}
S(q) = S_0(q) + \frac{1}{N} S_1(q) + \frac{1}{N^2}S_2(q) + ... ,
\end{align}
so that
\begin{align}\label{eq_vpm1}
r\left(q \pm \frac{1}{N}\right) &\propto e^{\pm S_0'\left(q\right) + \frac{1}{2N}S_0''(q) \pm \frac{1}{N}S_1'(q) } \nonumber \\
& \simeq e^{\pm S_0'\left(q\right)} \left[1+\frac{\frac{1}{2} S_0''(q) \pm S_1'(q)}{N}\right] .
\end{align}
Now we make the ansatz and obtain
\begin{alignat}{3}\label{eq_EVafterWKB}
&\quad &&e^{-S_0'\left(q\right)}\left[Nq^2 - q + \frac{q^2}{2}S_0''(q) - q^2S_1'(q) \right] \nonumber \\
& + &&e^{S_0'\left(q\right)}\left[Nq + \frac{q}{2}S_0''(q) + qS_1'(q)\right] \nonumber \\
&- &&\left(Nq + Nq^2 - q - qk_1\right) = N\xi_0 + \xi_1 .
\end{alignat}
In the $O(N)$ order we obtain a time-independent Hamilton-Jacobi equation
\begin{align}\label{eq_HJ_fromDV}
H_0\left[q,S_0'\left(q\right)\right] = \xi_0
\end{align}
with the Hamiltonian $H_0\left(q,p_0\right)$, defined in Eq.~\eqref{H0Ato2A}. The quantity $\xi_0$ plays the role of the (conserved) energy. We immediately notice that $H_0\left(q,p_0\right)$, and therefore $\xi_0$, do not depend on $k$, so that we can evaluate them at $k=0$.   But for $k=0$, the right eigenvector, which corresponds to the maximal eigenvalue, is a constant which can be set to be $r_{k=0} = 1$ \cite{Touchette2018}. As a result, $p_0(q) \equiv S_0'\left(q\right)=0$, and Eq.~\eqref{eq_HJ_fromDV} yields $\xi_0 = \xi_0^{(k=0)} = 0$.  This result coincides with the one obtained by the direct WKB method, where we found that the energy of the optimal trajectory is zero for any $0< a\leq 1$ \cite{fn_eigenvector}.

The sub-leading correction emerges from the terms $O(1)$ of Eq.~\eqref{eq_EVafterWKB}. Using the equalities $p_0(q) = p_0'(q) = 0$, we obtain
\begin{equation}\label{eq_xi1}
q(1-q)S'_1(q) + k_1q =\xi_1.
\end{equation}
This is again a time-independent Hamilton-Jacobi equation, with the Hamiltonian
$H_{1}\left(q,p_{1}\right)=q\left(1-q\right)p_{1}+k_{1}q$. Solving Eq.~(\ref{eq_xi1}) for $S_1'(q)$, we see that $S_1'(q)$ diverges at $q=1$ unless we set $\xi_1=k_1$.
As a result, $S_1(q)= k_1\ln q$ \cite{fn_subleading}. The divergence of $S_1(q)$ at $q=0$ is not dangerous, because at $q=O(1/N)$, that is at $n=O(1)$, our continuous WKB approximation for $r_n$ breaks down anyway.

By virtue of $\xi_1=k_1$ and Eqs. \eqref{eq_SCGF_byDV} and \eqref{eq_lambda_expansionToSecondOrder}, we obtain
\begin{equation}\label{eq_lambda1_xi1_k1}
\Lambda\left(k_1\right) = k_1 .
\end{equation}
Before we use this result in Eq.~\eqref{eq_LFtransform_secondOrder}, we will show that $k_1$ is bounded from below by $-1$.
Let us return to the exact eigenvalue problem~(\ref{eq_eigenvaluesProblem}):
\begin{equation}\label{eq_rightRecRel}
\frac{n\left(n-1\right)}{N}r_{n-1} + nr_{n+1} - \left[n + \frac{n\left(n-1\right)}{N} - nk\right]r_n = \xi r_n .
\end{equation}
In particular, for $n=1$ we obtain
\begin{equation}
\label{ratio}
\frac{r_2}{r_1} = 1-k+\xi .
\end{equation}
All components of the eigenvector, corresponding to the maximum eigenvalue, and in particular $r_1$ and $r_2$, must have the same sign \cite{Touchette2015,Touchette2018}. (A familiar analog is a ground-state wave function which does not have nodes.)
Therefore, the right hand side of Eq.~(\ref{ratio}) must be positive.  Using the scalings $k \simeq k_1/N$ and $\xi \simeq \xi_1$ [see Eqs.~\eqref{eq_SCGF_byDV}, \eqref{eq_k_expansionToSecondOrder} and \eqref{eq_lambda_expansionToSecondOrder}], we obtain
\begin{align}
1-\frac{k_1}{N}+\xi_1 > 0.
\end{align}
To avoid excess of accuracy, we ignore the term $-k_1/N$ and obtain the inequality $\xi_1>-1$. Applying now the WKB result \eqref{eq_lambda1_xi1_k1}, we see that $k_1>-1$.
To remind the reader, this result is valid only in the WKB regime. Using this bound and Eq.~\eqref{eq_lambda1_xi1_k1} in Eq.~\eqref{eq_LFtransform_secondOrder}, we obtain
\begin{align}
I_1(a < 1) = \max_{k_1>-1}\{k_1a - k_1\} = 1-a,
\label{aless1}
\end{align}
that is the maximum is reached at the boundary. The nonzero rate function (\ref{aless1}) removes the degeneracy $I\left(a,N\right)=0$ at $0\leq a\leq 1$,
that we previously observed at $N\to \infty$.

The continuous WKB theory breaks down at $a=O\left(1/N\right)$, so Eq.~(\ref{aless1}) is inapplicable there. It is also inapplicable too close to $a=1$, where the WKB action is small. Fortunately, in a close  vicinity of  $a=1$ one can approximate \cite{AssafMeerson,MeersonZilber} the exact master equation of the jump process by a continuous Fokker-Planck equation, using the Van Kampen system-size expansion \cite{Kampen}. This leads to an effective Ornstein-Uhlenbeck process, and we obtain (see \textit{e.g.} Eq. (49) in Ref. \cite{Touchette2018})
\begin{equation}\label{eq_FPapproximation}
I\left(a,N\right)\simeq \frac{N}{4}\left(a-1\right)^{2},
\end{equation}
corresponding to the Gaussian asymptotic of $\mathcal{P}\left(\bar{n}_{T}=a\right)$  as a function of $a$ near the expected value $a=1$. Equation~(\ref{eq_FPapproximation}) also follows from a Taylor expansion around $a=1$ of the $a>1$ branch of $I(a)$ from Eq.~\eqref{eq_A2A2AA_correctf}. Notably, the smooth Gaussian asymptotic clearly implies that the second-order phase transition
is an attribute of the leading-order WKB approximation. Using Eqs.~(\ref{aless1}) and (\ref{eq_FPapproximation}), we can establish the applicability condition of Eq.~(\ref{aless1}) at $a$ close to $1$: $1-a \gg 1/N$.
Putting everything together, we obtain
\begin{numcases}
{\!\!\!\!\frac{I\left(a,N\right)}{N}  \simeq \!}
\!\frac{1-a}{N}, & $a \! \gg \! \frac{1}{N}$ \text{and} $1\!-\!a \!\gg \! \frac{1}{N}$, \label{theory1}\\
\!\frac{\left(1-a\right)^{2}}{4}, & $0\leq 1-a \ll \frac{1}{N}$  \nonumber \\
 & \text{or } $0\leq a-1\ll 1$, \label{theory2} \\
\!\left(a-\sqrt{a}\right)^{2}\!\!, & $a\geq 1$ \label{theory3}.
\end{numcases}
The asymptotics~(\ref{theory2}) and (\ref{theory3}) coincide in their joint region $0\leq a-1\ll 1$.

\begin{figure}[h]
\includegraphics[width=0.38\textwidth,clip=]{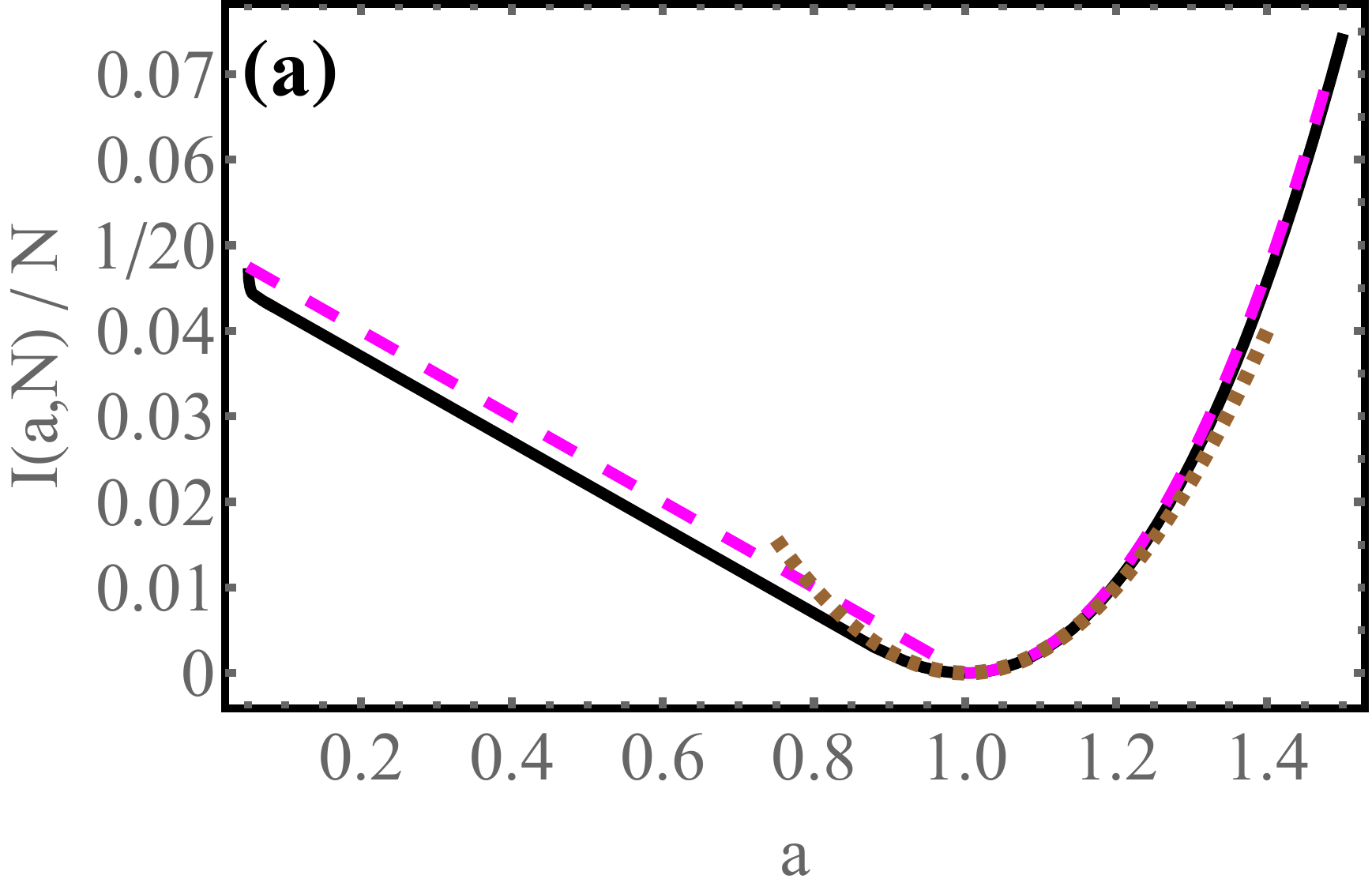}
\includegraphics[width=0.38\textwidth,clip=]{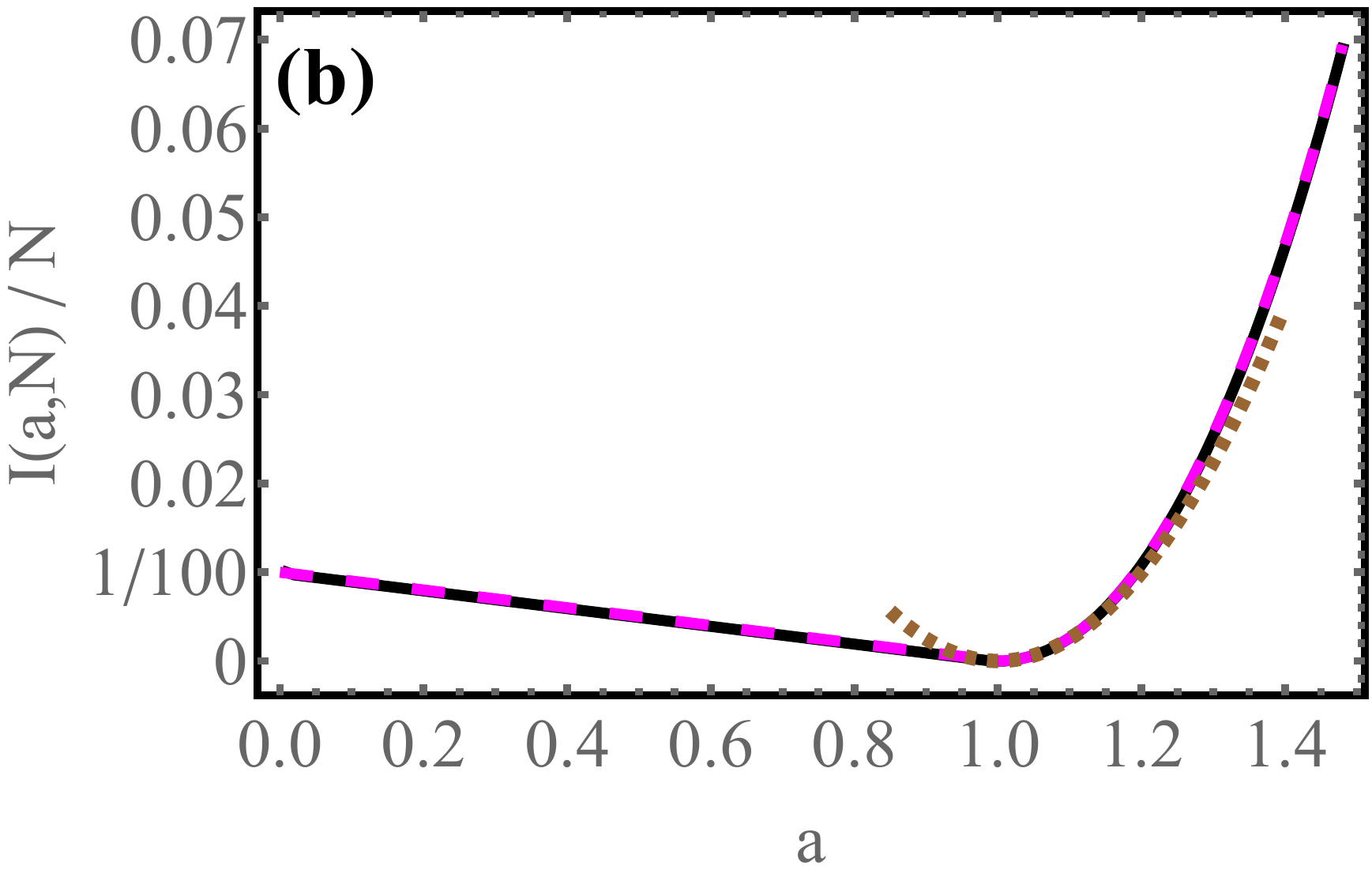}
\caption{The rescaled rate function $I\left(a,N\right)/N$ for the process $A\rightleftarrows 2A$ with  $N=20$ (a) and $N=100$ (b), as predicted by WKB theory  [Eqs.~\eqref{theory1} (dashed line), \eqref{theory2} (dotted line) and
\eqref{theory3} (dashed line)] is compared with numerical results, obtained by the DV method \cite{DonskerVaradhan} (solid line). The WKB approximation improves for larger $N$. Noticeable in (a) is an infinite negative slope at $a=1/N$.}
\label{fig_A2A2AA_WKB1}
\end{figure}

Our numerics, see Fig.~\ref{fig_A2A2AA_WKB1}, show that the exact rate function is strictly convex. This feature is most clearly observed at not too large $N$ or at small $a$. The approximate expression (\ref{theory1})  is not strictly convex, but it is natural to assume that strict convexity will appear in the next order of the perturbation procedure \cite{minus_infinity}. Figure~\ref{fig_A2A2AA_WKB1}  compares the  asymptotics (\ref{theory1})-(\ref{theory3}) with numerical results, obtained by the DV method \cite{DonskerVaradhan}, for $N=20$ and $100$. As one can see, the WKB approximation improves for larger $N$.

\begin{figure}[h]
\includegraphics[width=0.38\textwidth,clip=]{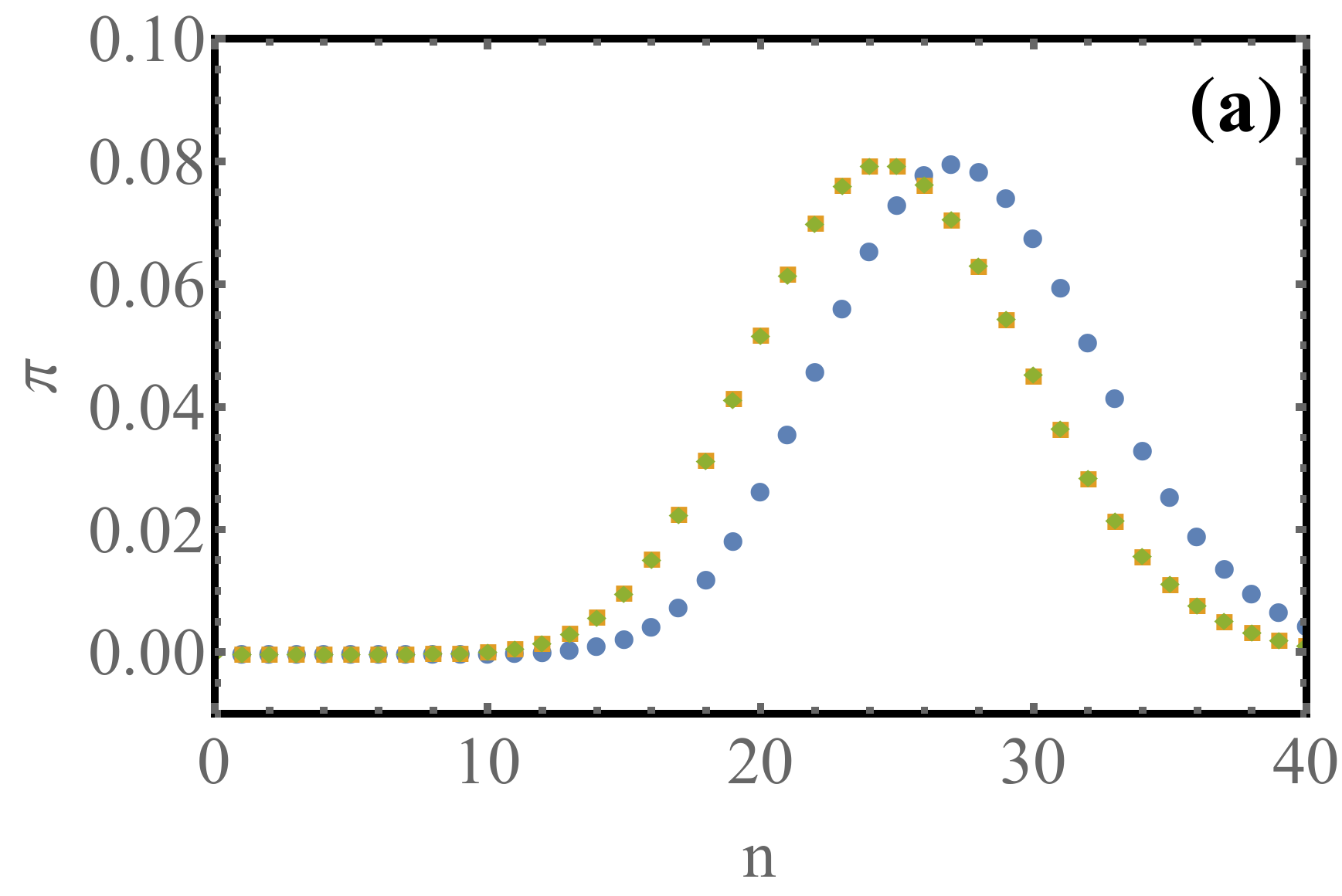}
\includegraphics[width=0.38\textwidth,clip=]{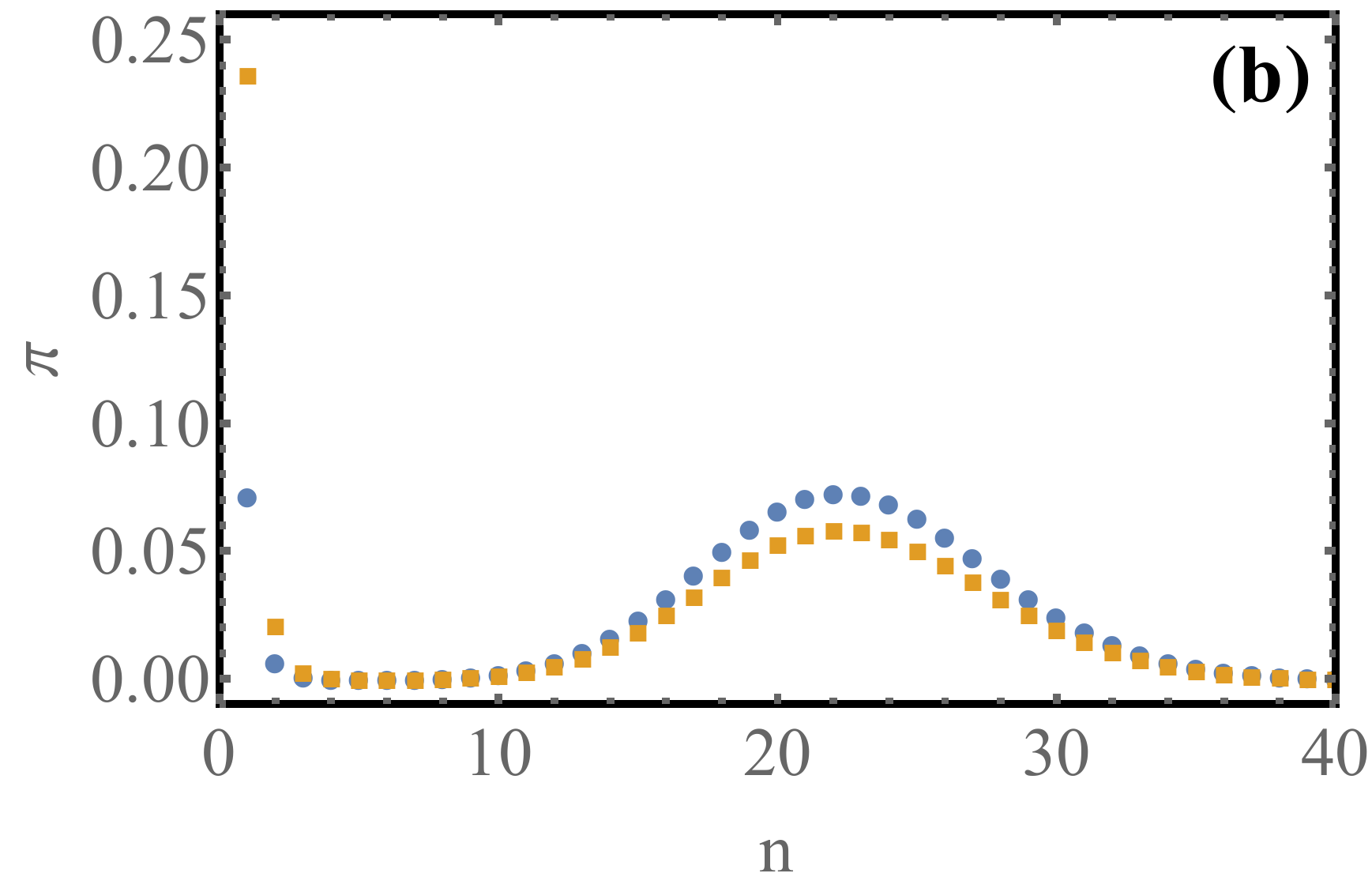}
\includegraphics[width=0.38\textwidth,clip=]{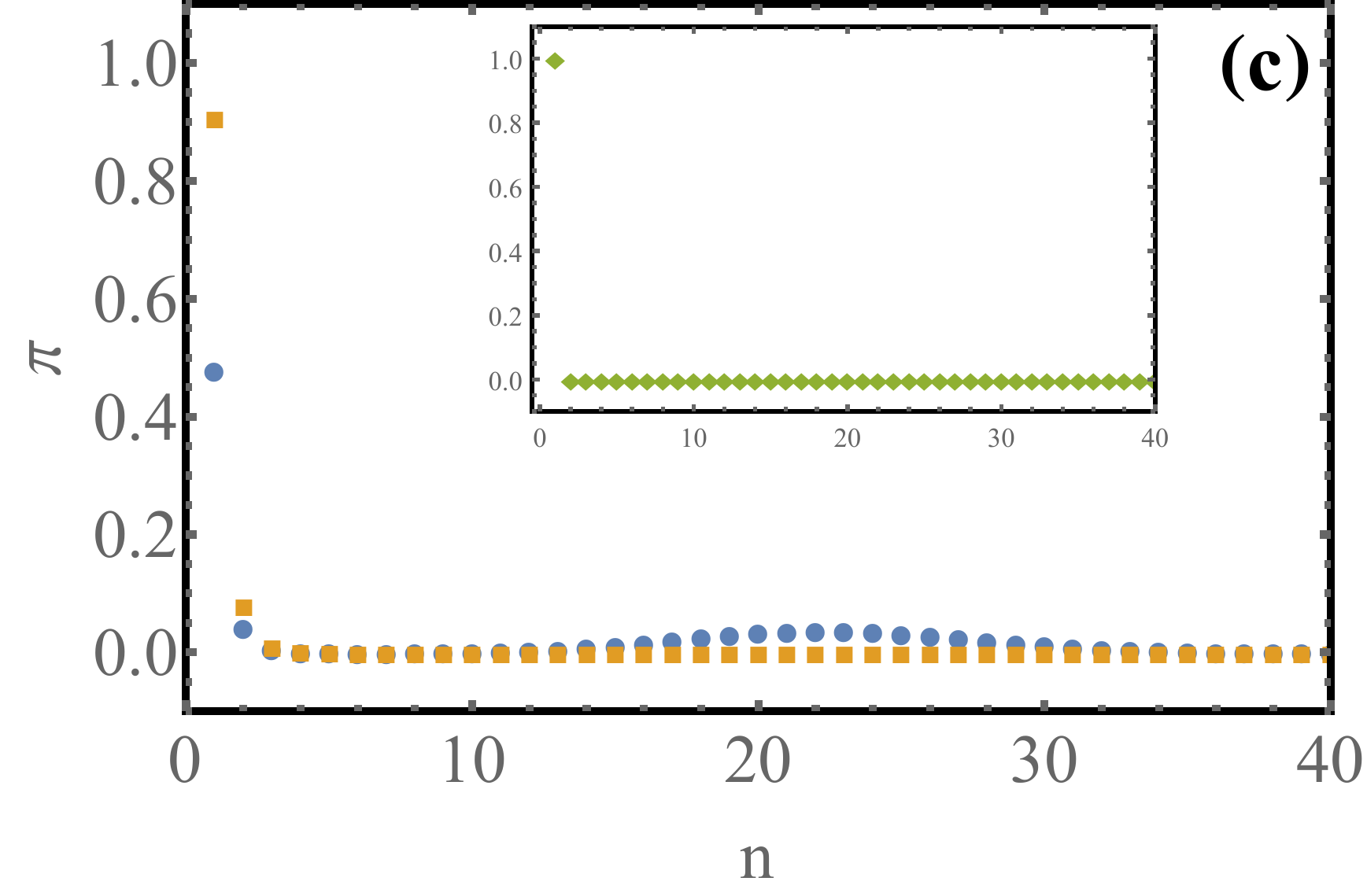}
\caption{Stationary distribution $\pi_n(a)$, conditional on a specified $a$, found numerically  by the DV method with $N=25$. (a): $a\simeq 1.10$ (circles) and $a\simeq 1.00$  (squares). Also shown by diamonds is the exact stationary distribution~(\ref{Poisson2}); the squares and diamonds coincide. (b): $a\simeq 0.84$ (circles) and  $a\simeq 0.68$ (squares). (c): $a\simeq 0.45$ (circles) and $a \simeq 0.045$ (squares). Inset: $a = 1/N =0.04$.
Note the different vertical scale in different panels. For $a<1$, $\pi$ is bimodal.}
\label{fig_distributions}
\end{figure}

Figure  \ref{fig_distributions} shows our numerical results for the stationary distribution $\pi_n(a)$ of the branching-annihilation process,  conditioned on a given value of $a$ (or $k$). This distribution is given by the properly normalized product of the right and left eigenvectors, $\pi_n = r_n l_n$ for a fixed value of $a$ (or $k$) \cite{Chetrite2013,Touchette2015}.
Remarkably (but \textit{a posteriori} not surprisingly), for $1/N<a<1$ the conditional distribution  $\pi_n$ is bimodal: one of its two peaks is  at $n\simeq N$, the other is at $n=1$.  As $N$ increases, the two peaks become well separated, and the area under the peak at $n=N$ approaches $a$, whereas the area under the peak at $n=1$ approaches $1-a$. This is in a remarkable agreement with the prediction of our direct WKB analysis of Sec. \ref{sec_WKB0_examples}. Indeed, for $a<1$ the optimal trajectory is such that the population has size $N$ for time $aT$ and then, after a short transient, size $1/N$ (which, in the direct WKB method, is indistinguishable from zero) for time $(1-a)T$.

\section{Discussion}
\label{sec_discussion}

Time-averaged quantities provide a useful characterization of stochastic populations. Here, using a WKB approximation, we observed, at $N\to \infty$, a dynamical phase transition and a giant disparity in the large deviation function (LDF) of the time-averaged population size of the branching-coalescence process. These effects result from a qualitative change in the ``optimal" trajectory of the underlying Hamiltonian classical mechanics problem. The phase transition is in fact smoothed at finite $N$, but the probabilities of observing unusually small and unusually large values of the time-averaged population size remain dramatically different. In order to resolve the degeneracy $f(a)=0$ at $0\leq a\leq 1$, obtained in the leading WKB order, we developed a different version of WKB method, which is applied in conjunction with the Donsker-Varadhan large-deviation formalism and goes beyond the leading-order calculations in $1/N$. A similar in spirit methodology has been recently implemented for a continuous diffusion process \cite{Derrida}, where a (very different) singularity of a large deviation function, and its regularization, have been studied.

From a more general perspective, our work provides an additional example of a singularity in a LDF of a time-integrated fluctuating quantity which appears in the weak-noise limit $N\to \infty$. A somewhat similar in spirit singularity  (also of second order) appears in the LDF of the time- and space-integrated current in certain diffusive lattice gases in systems with periodic boundaries \cite{Bertini2006,Bodineau2007}. The details, however, are quite different.

Our results, obtained for the branching-annihilation model, can be immediately  generalized to a whole family of single-species stochastic population models, which have a nontrivial (that is, non-empty) steady state, corresponding to a non-trivial attracting fixed point of deterministic theory. A sufficient condition for the emergence of a giant probability disparity at large $N$, and a second-order dynamical phase transition in the limit of $N\to\infty$, is the presence of a repelling fixed point at $q=0$. We can even relax the demand of the non-empty steady state by conditioning the process on the population survival by the time $T$ over which the long-time average is calculated.

It would be interesting to complete our calculations of $I\left(a,N\right)$ in the regions  $a=O\left(1/N\right)$ and $1-a=O\left(1/N\right)$, where WKB theory breaks down.  It would be also interesting to study statistics of the time-averaged population size for stochastic populations which exhibit switches between two metastable states. In the deterministic description, these populations have two attracting fixed points, and a repelling point in between. A host of additional interesting questions concerns multi-population systems which describe coexistence of species.

Finally,  a giant disparity and a second-order dynamical phase transition should also appear,  in the weak-noise limit,  in LDFs of time-averaged quantities in continuous systems with multiplicative noise. In analogy to our work, this can happen in the presence of an attracting fixed point $q\neq 0$ and a repelling fixed point $q=0$ in the deterministic limit.

\section*{ACKNOWLEDGMENTS}
We are grateful to Hugo Touchette and Tal Agranov for useful discussions, and acknowledge support from the Israel Science Foundation (Grant No. 807/16). N.R.S. was supported by the Clore Foundation.


\begin{thebibliography} {99}
\bibitem{fn_nmin} As the state $n=0$ is inaccessible, the state $n=1$ is the best proxy, at large $N$, of the ``classical" fixed point $q=0$.
\bibitem{lessinteresting} For $a>1$ we calculated the subleading term $I_1(a)$ in Eq.~(\ref{eq_IexpansionToSecondOrder_full}) and verified it against numerical calculations. We do not present these results here.
\bibitem{fn_eigenvector} When using the direct WKB method in Sec. \ref{sec_WKB0_formalism}, we worked with the left eigenvector, which corresponds to the thick line in Fig.~\ref{constrainedH}b. Now we are working with the right eigenvector, which is described by the deterministic line $p=0$ on Fig.~\ref{constrainedH}b. Both vectors are described by the zero-energy trajectories of the Hamiltonian~\eqref{H0Ato2A}.
\bibitem{fn_subleading} $S_1(q)$ yields the subleading correction to the right eigenvector, which can be used when computing sub-subleading corrections.
\bibitem{minus_infinity} We proved that the exact rate function $I\left(a,N\right)$ exhibits an infinite negative slope as a function of $a$ at the point $a_{\text{min}}=1/N$ (that is, beyond the applicability domain of the WKB theory).
This feature, and the Fokker-Planck result \eqref{theory2} for $a\simeq 1$, provide analytic arguments in favor of the strict convexity of the exact rate function.
\end{thebibliography}
\end{document}